# Padded Helmet Shell Covers in American Football: A Comprehensive Laboratory Evaluation with Preliminary On-Field Findings

Nicholas J. Cecchi[1]*, Ashlyn A. Callan[1]*, Landon P. Watson[1], Yuzhe Liu[1], Xianghao Zhan[1], Ramanand V. Vegesna[2], Collin Pang[1], Enora Le Flao[1], Gerald A. Grant[3,4,5], Michael M. Zeineh[6], David B. Camarillo[1,3,7]

1. Department of Bioengineering, Stanford University, Stanford, CA 94305, USA
2. Department of Biomedical Engineering, University of Southern California, Los Angeles, CA 90089, USA
3. Department of Neurosurgery, Stanford University, Stanford, CA 94305, USA
4. Department of Neurology, Stanford University, Stanford, CA 94305, USA
5. Department of Neurosurgery, Duke University, NC 27710, USA
6. Department of Radiology, Stanford University, Stanford, CA 94305, USA
7. Department of Mechanical Engineering, Stanford University, CA 94305, USA
*co-first author

**Abstract:**

Protective headgear effects measured in the laboratory may not always translate to the field. In this study, we evaluated the impact attenuation capabilities of a commercially available padded helmet shell cover in the laboratory and field. In the laboratory, we evaluated the padded helmet shell cover's efficacy in attenuating impact magnitude across six impact locations and three impact velocities when equipped to three different helmet models. In a preliminary on-field investigation, we used instrumented mouthguards to monitor head impact magnitude in collegiate linebackers during practice sessions while not wearing the padded helmet shell covers (i.e., bare helmets) for one season and whilst wearing the padded helmet shell covers for another season. The addition of the padded helmet shell cover was effective in attenuating the magnitude of angular head accelerations and two brain injury risk metrics (DAMAGE, HARM) across most laboratory impact conditions, but did not significantly attenuate linear head accelerations for all helmets. Overall, HARM values were reduced in laboratory impact tests by an average of 25% at 3.5 m/s (range: 9.7 – 39.6%), 18% at 5.5 m/s (range: -5.5% – 40.5), and 10% at 7.4 m/s (range: -6.0 – 31.0%). However, on the field, no significant differences in any measure of head impact magnitude were observed between the bare helmet impacts and padded helmet impacts. Further laboratory tests were conducted to evaluate the ability of the padded helmet shell cover to maintain its performance after exposure to repeated, successive impacts and across a range of temperatures. This research provides a detailed assessment of padded helmet shell covers and supports the continuation of in vivo helmet research to validate laboratory testing results.

**Key Terms:** Concussion, brain strain, instrumented mouthguard, head kinematics, headgear, brain injury, temperature

**Introduction:**

Repeated head impact exposure and concussions are frequent in contact sports, particularly in American football.[12,29] This trauma to the brain has been associated with acute and long-term deterioration of brain health and function, including negative physiological and psychological effects in athletes.[1,28,33] To mitigate the risk of injuries to the head and brain, protective helmets have been worn in American football for over a century. Since their initial implementation in the sport, helmets have seen significant changes in their appearance, size, weight, material composition, and efficacy in attenuating the severity of head impacts.[46] As awareness and concern surrounding sport-related brain safety continue to grow, research and development efforts in helmet technology have yielded a wide variety of novel approaches to mitigating impact energy.[2,11,45]

Presently, American football helmets are evaluated for their efficacy in mitigating the risk of serious head injury (i.e., skull fracture)[36] and concussion.[4,36,44] Head kinematics resulting from impacts to the helmet, and several brain injury metrics derived from these kinematic measurements, have been found to closely correlate with the risk of on-field concussive injury[19,38] and are therefore frequently used by researchers to describe the relative severity of impacts and performance of helmets. Nearly all quantitative studies describing helmet performance are conducted using laboratory impact testing equipment (drop towers, pneumatic linear impactors, pendulum impactors, etc.) and anthropomorphic test devices created to replicate human head impact dynamics.[4,44,47]

On-field studies of head impacts have been made possible through use of a variety of head impact sensor technologies.[25] Instrumented mouthguards have been found to provide considerably accurate measurement of head impact kinematics resulting from sport-related impacts due to their firm connection to the upper dentition of the maxilla, rigidly connected to the skull.[27] These mouthguard-measured kinematics have further been utilized to estimate brain strains and various brain injury criteria in athlete populations.[26,27,43,48] Advances in instrumented mouthguard technology enable researchers to determine the effect of various sport-specific factors on impact frequency and severity,[9,30] though limited studies utilizing head impact sensors have investigated the effect of helmet technology on impact severity in on-field, human studies.[8,9,40]

One recent innovation in helmet technology is the introduction of padded shell covers that are made to fit over the exterior of a helmet.[2,7,49] This technology is intended to reduce the severity of head impacts by (i) adding additional, compressible padding to the outside of a helmet, (ii) lowering the surface friction of the helmet's exterior, and (iii) redirecting impact energy through independent rotation of the padded helmet shell cover along the surface of the helmet shell. These products have become commercially available for athletes and are rapidly gaining popularity, with some even becoming mandated under certain conditions at the elite level of American football competition.[35] Various previous studies of this form of helmet technology have been conducted,[2,7,49] but a thorough investigation exploring attenuation of linear and angular head kinematics and brain strain, robustness of attenuation after repeated impacts, impact response across varying temperatures, and on-field efficacy has yet to be conducted. Before widespread adoption of padded helmet shell covers, a full understanding of their safety performance is necessary. The objective of this study was to comprehensively evaluate the efficacy of a commercially available padded helmet shell cover in attenuating linear and angular head kinematics and brain strain under various laboratory impact conditions. Further, we sought to gather preliminary data that could aid in determining if our laboratory findings translated to an on-field setting.

**Methods:**

*Laboratory Helmet Impact Testing*

Laboratory impact tests were conducted on three helmet models with and without a padded helmet shell cover equipped to each. The three helmet models were the Riddell Speedflex Diamond (Riddell Sports Group, Inc.; Rosemont, IL), the Xenith X2E+ (Xenith, LLC; Detroit, MI), and the Schutt Vengeance Pro LTD (Schutt Sports, Inc.; Litchfield, IL). These helmets were chosen to represent a range of performance rankings according to the 2021 National Football League (NFL) helmet rankings.[34] The padded helmet shell cover used was the Guardian Cap (Guardian Sports; Peachtree Corners, GA), a one-size-fits-all padded helmet shell cover composed of a closed-cell polyurethane foam. For each helmet, the padded helmet shell cover was attached to the helmets by self-attaching elastic straps and buckles that fit around the facemask of the helmet and an elastic and Velcro strap at the back of the helmet.

A pneumatic linear impactor (Biokinetics and Associates Ltd.; Ottawa, Ontario, Canada) was used to impact a helmeted 50th percentile male Hybrid III headform mounted on a sliding table for all helmet impact tests. Each helmet with and without the padded helmet shell cover equipped was impacted at three impact velocities (3.5, 5.5, and 7.4 m/s) representing a range of subconcussive and concussive impacts.[4] Six impact locations were tested (Fig. 1): four locations were entirely covered by the padded helmet shell cover (Side, Oblique Rear, Side Upper, Oblique Front), while two locations were located on the facemask (Facemask Side, Facemask Central Oblique). The six impact locations are described in further detail by Bailey et al.[4] During impact tests when the padded helmet shell cover was equipped to a helmet, an identical padded helmet shell cover was equipped to the end cap of the linear impactor. Three repeat impacts per helmet-location-velocity combination were performed. For each different helmet model, a new padded helmet shell cover was used for the helmet and the impactor. All impacts were within +/- 2% of the target impact velocity, as measured by a velocity gate on the linear impactor.

A separate set of impact tests was conducted to evaluate the robustness of the padded helmet shell cover after sustaining multiple impacts. Considering that frontal impacts have been found to be the most frequent among football players at various skill levels,[12] 100 consecutive impacts were delivered to the oblique front location at 3.5 m/s while a Riddell Speedflex Diamond helmet was fit to the headform and a padded helmet shell cover was equipped to both the helmet and the end cap of the linear impactor. The delay between each impact was approximately 80 seconds.

For all helmet shell impacts, the headform was equipped with a six-degree-of-freedom sensor package including three DTS A64C accelerometers (Diversified Technical Systems, Inc.; Seal Beach, CA) and three DTS ARS PRO-8k gyroscopes. Facemask impacts were performed at a later date than the shell impacts, when the headform was equipped with a six-degree-of-freedom sensor package including a DTS 6DX Pro sensor. For all impacts, linear acceleration and angular velocity in the X, Y, and Z directions was captured at 20,000 Hz using DTS SLICEWare software from a SLICE MICRO data acquisition system.

Linear accelerations and angular velocities of verified laboratory impacts were filtered by a fourth-order Butterworth filter at 300 Hz. A five-point stencil derivative was performed on the angular velocity measurements to calculate angular accelerations, which were also filtered by a fourth-order Butterworth filter at 300 Hz. Peak linear acceleration (PLA) and peak angular acceleration (PAA) were extracted for each processed impact event. To estimate the maximum brain strain resulting from each impact, the Diffuse Axonal Multi-Axis General Evaluation (DAMAGE) metric was calculated according to the equations described by Gabler et al.[18] The Head Acceleration Response Metric (HARM) was also calculated for each impact according to the equations described by Bailey et al.[4]

*Vertical Drop Testing*

Further testing using a twin-wire drop apparatus was conducted to determine the effect of repeated impacts and changes in environmental temperature on the impact response of the padded helmet shell cover. Nine identical sample sections were cut out from the crown of padded helmet shell covers. Each sample was fixed to a force plate (430_00_LCEL; Cadex Inc; St-Jean-sur-Richelieu, Canada) and impacted at 2.0 m/s by a 2.5kg dropping mass guided by twin wires (Fig. 2). Three samples were tested at ambient temperature (21ºC), three samples were tested within 90 seconds of 18-hour exposure to a cold environment (a freezer set to -18ºC), and three samples were tested within 90 seconds of 18-hour exposure to a hot environment (an incubator set to 50ºC). At the ambient condition, samples were impacted three consecutive times, with a delay of 80 seconds between impacts. Testing for each sample was repeated twice (after a delay of 24 hours for the padding to recover), totaling six trials for each condition. The force measurements collected by each of the three load cells in the force plate were independently filtered by a 2nd-order low-pass Butterworth filter with a cut-off frequency of 300 Hz, and then summed as the reaction force of the padding sample. The peak force was recorded for each impact.

*On-field Data Collection*

As part of a larger study, athletes from the Stanford University football team were equipped with MiG2.0 instrumented mouthguards (Stanford University; Stanford, CA) during official team practices in two separate years of training. In the first year (2019), the team wore bare helmets without the padded helmet shell cover; in the second year (2021), coaching staff required athletes to wear the padded helmet shell covers. The team wore padded helmet shell covers during parts of 2020, but data collection did not take place that year due to the COVID-19 pandemic. Five linebackers were recruited during each year of data collection, with data being collected from 13 practices in the first year and 14 practices in the second year. All activities involving human subjects were approved by the Institutional Review Board at Stanford University.

The MiG2.0 mouthguards were custom fit to each athlete's upper dentition and acquired linear accelerations and angular velocities of head impact events via a triaxial accelerometer and triaxial gyroscope, respectively. The triaxial accelerometer recorded linear acceleration data at

1000 Hz and the triaxial gyroscope recorded angular velocity data at 8000 Hz. The mouthguards were set to record a 200 ms time window of data upon any direction of linear acceleration exceeding a 10 g threshold (50 ms pre-trigger, 150 ms post-trigger). Previous studies have found that MiG2.0 devices operating with these settings are capable of accurately measuring linear and angular head kinematics for use as input to various brain injury metrics.[26,27]

Events recorded by the mouthguards were then subjected to two steps of verification. First, the events were processed through the MiGNet program. MiGNet is a deep learning algorithm that was previously validated to distinguish true head impacts from false positive events caused by application, removal, or handling of the devices or other actions unrelated to impacts.[17] Second, the events indicated as true positives by the MiGNet program were visually verified using video footage of practices. Practice footage was filmed at 1080p resolution and 60 frames per second by videographers at four camera angles simultaneously, encompassing the majority of the practice field. Video footage only captured intermittent periods of active drills and practice time and did not capture breaks in between drills. A world clock was made visible during filming to enable temporal synchronization between the captured mouthguard data and practice footage. Only helmet-to-helmet impacts were retained for analyses, rather than impacts caused by body-to-body, body-to-head, or head-to-ground contact, to best compare the bare and padded helmet impacts (Fig. 3). A helmet-to-helmet impact for the padded condition was classified as any impact where the helmet of one player equipped with a padded helmet shell cover hit the helmet of another player equipped with a padded helmet shell cover. If only one player was wearing a padded helmet shell cover, the impact was discarded and not included in any analyses. A bare helmet impact was classified as any impact where the helmet of one player hit the helmet of another player, where neither player was fitted with a padded helmet shell cover. Impacts that did not involve helmet-to-helmet contact were discarded. Two of the authors, who each had two years of experience reviewing sport-related head impact events, independently analyzed each impact. In cases where the two video reviewers reached inconsistent classifications, a third author with seven years of experience reviewing sport-related head impact events independently reviewed the impact to make a final decision on impact validity. Confirmed impacts were then further classified as either "shell" or "facemask" impacts. "Shell" impacts were defined as those in which the primary point of contact for at least one player was located on the helmet shell. "Facemask" impacts were those in which the facemask was identified as a primary point of contact for both players involved in the impact.

Linear accelerations and angular velocities of verified on-field impact events were filtered by a fourth-order Butterworth filter at 160 Hz. A five-point stencil derivative was performed on the angular velocity measurements to calculate angular accelerations, which were also filtered by a fourth-order Butterworth filter at 160 Hz. Linear accelerations were transformed from the location of the accelerometer to the estimated center of gravity of the head. PLA, PAA, DAMAGE, and HARM were calculated for each processed impact event.

*Statistical Analyses*

All statistical analyses were performed in Prism 9.0.1 (Graphpad Software, Inc; San Diego, CA). For laboratory impact attenuation tests at helmet shell locations, twelve two-way ANOVAs were performed to analyze the effect of impact condition (i.e., location-velocity combination) and use of padded helmet shell cover on head impact response (3 independent helmets: Riddell, Xenith, Schutt; 4 independent head response metrics: PLA, PAA, DAMAGE, HARM). For laboratory impact attenuation tests at facemask locations, twelve two-way ANOVAs were performed to analyze the effect of impact condition (i.e., location-velocity combination) and use of padded helmet shell cover on head impact response (3 independent helmets: Riddell, Xenith, Schutt; 4 independent head response metrics: PLA, PAA, DAMAGE, HARM). For repeated drop impact tests, a one-way ANOVA was performed to compare the effect of impact count (one, two, and three consecutive impacts) on peak force upon impact. A simple linear regression was used to test if impact count significantly affected the magnitude of HARM (i.e., whether the regression slope significantly deviated from zero) across 100 consecutive impacts in full helmet repeated impact tests. A one-way ANOVA was performed to compare the effect of temperature (cold, ambient, and hot) on peak force upon impact. For our on-field dataset, separate Mann-Whitney tests were used to test for differences in PLA, PAA, DAMAGE, and HARM for impacts sustained between bare helmets and helmets equipped with the padded helmet shell covers.

## **Results:**

*Laboratory Impact Attenuation*

The effect of the padded helmet shell covers on peak head kinematics, DAMAGE, and HARM varied by impact condition and helmet model (Fig. 4). For PLA at shell impact locations, two-way ANOVAs revealed that there was a statistically significant interaction between the effects of impact condition and padded helmet shell cover use in the Riddell ($F(11,48) = 2.390$, $p = 0.019$), Xenith ($F(11,48) = 4.141$, $p = 0.0003$), and Schutt ($F(11,48) = 9.242$, $p < 0.0001$) helmets. Simple main effects analysis showed that impact condition had a significant effect on PLA at shell impact locations for all three helmets ($p < 0.0001$) and that padded helmet shell cover use had a significant effect on PLA in the Xenith helmet ($p < 0.0001$), but not the Riddell ($p = 0.307$) or Schutt ($p = 0.097$) helmet. At facemask impact locations, two-way ANOVAs revealed that there was not a statistically significant interaction between the effects of impact condition and padded helmet shell cover use in the Riddell ($F(5,24) = 1.173$, $p = 0.351$), Xenith ($F(5,24) = 1.460$, $p = 0.240$), and Schutt ($F(5,24) = 0.5181$, $p < 0.760$) helmets. Simple main effects analysis showed that impact condition had a significant effect on PLA at facemask impact locations for all three helmets ($p < 0.0001$) and that padded helmet shell cover use had a significant effect on PLA in the Xenith helmet ($p = 0.0001$), but not the Riddell ($p = 0.783$) or Schutt ($p = 0.530$) helmets.

For PAA at shell impact locations, two-way ANOVAs revealed that there was a statistically significant interaction between the effects of impact condition and padded helmet shell cover use in the Riddell ($F(11,48) = 29.05$, $p < 0.0001$), Xenith ($F(11,48) = 24.11$, $p < 0.0001$), and Schutt ($F(11,48) = 6.745$, $p < 0.0001$) helmets. Simple main effects analysis showed that impact condition had a significant effect on PAA at shell impact locations for all three helmets ($p < 0.0001$) and that padded helmet shell cover use had a significant effect on PAA for all three helmets ($p < 0.0001$). At facemask impact locations, two-way ANOVAs revealed that there was a statistically significant interaction between the effects of impact condition and padded helmet shell cover use in the Riddell ($F(5,24) = 13.12$, $p < 0.0001$) and Xenith ($F(5,24) = 5.343$, $p = 0.002$) helmets, but not the Schutt ($F(5,24) = 0.6631$, $p = 0.655$) helmet. Simple main effects analysis showed that impact condition had a significant effect on PAA at facemask impact locations for all three helmets (p < 0.0001) and that padded helmet shell cover use had a significant effect on PAA in the Riddell (p = 0.035), Xenith (p = 0.004), and Schutt (p < 0.0001) helmets.

For DAMAGE at shell impact locations, two-way ANOVAs revealed that there was a statistically significant interaction between the effects of impact condition and padded helmet shell cover use in the Riddell ($F_{(11,48)}$ = 17.58, p < 0.0001), Xenith ($F_{(11,48)}$ = 43.14, p < 0.0001), and Schutt ($F_{(11,48)}$ = 12.76, p < 0.0001) helmets. Simple main effects analysis showed that impact condition had a significant effect on DAMAGE at shell impact locations for all three helmets (p < 0.0001) and that padded helmet shell cover use had a significant effect on DAMAGE for all three helmets (p < 0.0001). At facemask impact locations, two-way ANOVAs revealed that there was a statistically significant interaction between the effects of impact condition and padded helmet shell cover use in the Riddell ($F_{(5,24)}$ = 26.59, p < 0.0001) helmet, but not the Xenith ($F_{(5,24)}$ = 0.8840, p = 0.507) or Schutt ($F_{(5,24)}$ = 0.8848, p = 0.506) helmets. Simple main effects analysis showed that impact condition had a significant effect on DAMAGE at facemask impact locations for all three helmets (p < 0.0001) and that padded helmet shell cover use had a significant effect on DAMAGE in the Riddell (p < 0.0001), Xenith (p < 0.0001), and Schutt (p = 0.0001) helmets.

For HARM at shell impact locations, two-way ANOVAs revealed that there was a statistically significant interaction between the effects of impact condition and padded helmet shell cover use in the Riddell ($F_{(11,48)}$ = 6.252, p < 0.0001), Xenith ($F_{(11,48)}$ = 36.84, p < 0.0001), and Schutt ($F_{(11,48)}$ = 9.618, p < 0.0001) helmets. Simple main effects analysis showed that impact condition had a significant effect on HARM at shell impact locations for all three helmets (p < 0.0001) and that padded helmet shell cover use had a significant effect on HARM for all three helmets (p < 0.0001). At facemask impact locations, two-way ANOVAs revealed that there was a statistically significant interaction between the effects of impact condition and padded helmet shell cover use in the Riddell ($F_{(5,24)}$ = 17.35, p < 0.0001) helmet, but not the Xenith ($F_{(5,24)}$ = 0.6835, p = 0.640) or Schutt ($F_{(5,24)}$ = 0.8819, p = 0.533) helmets. Simple main effects analysis showed that impact condition had a significant effect on HARM at facemask impact locations for all three helmets (p < 0.0001) and that padded helmet shell cover use had a significant effect on HARM in the Riddell, Xenith, and Schutt helmets (p < 0.0001).

In 33 of the 36 helmet shell test conditions in this study, the use of the padded helmet shell cover resulted in a lower average HARM value than the bare helmet tests. For all three helmet models, the percentage reduction in HARM at shell impact locations decreased as impact velocity increased. In all 18 facemask test conditions, the use of the padded helmet shell cover resulted in a lower average HARM value than the bare helmet tests. Across all impact locations and helmet models, HARM values were reduced in laboratory impact tests by an average of 25% at 3.5 m/s (range: 9.7 – 39.6%), 18% at 5.5 m/s (range: -5.5% – 40.5), and 10% at 7.4 m/s (range: -6.0 – 31.0%). Average percentage reduction in HARM for each individual shell impact condition can be found in Figure 5A. Average percentage reduction in HARM for each individual facemask impact condition can be found in Figure 5B.

*Repeated Impact Testing*

For linear drop testing on a cutout section of the padded helmet shell covers, one-way ANOVA results revealed that there was a statistically significant difference in peak force between at least two impact count groups ($F(2, 15) = 9.125$, $p = 0.003$). Tukey's multiple comparisons tests revealed that a 3% increase in mean peak force after two impacts compared to one impact ($p = 0.047$) and a 5% increase in mean peak force after three impacts compared to one impact ($p = 0.002$) were statistically significant, while differences in peak force between the second and third impacts ($p = 0.276$) were not significant (Fig. 6A). HARM values from full helmet repeated impacts ranged from 1.55 to 1.79; results of a simple linear regression showed that the regression line slope was not significantly different from zero and impact count was not associated with HARM magnitude ($R^2 = 0.04187$, $p = 0.830$) (Fig. 6B).

*Temperature Sensitivity*

One-way ANOVA results revealed that there was a statistically significant difference in peak force between at least two temperature groups ($F(2, 15) = 6.080$, $p = 0.012$). Tukey's multiple comparisons tests revealed the 13% difference between means in peak force at the cold and hot temperature conditions was statistically significant ($p = 0.011$), while differences in peak force between cold and ambient ($p = 0.072$) and ambient and hot ($p = 0.597$) did not meet statistical significance (Fig. 7).

*On-Field Impact Attenuation*

Instrumented mouthguards were worn for a total of 43 athlete-exposures in 2019 and 58 athlete exposures in 2021. After MiGNet processing and video review, 97 events were deemed to be helmet-to-helmet impacts by the video reviewers, including 46 bare helmet impacts and 51 padded helmet impacts. Both players' facemasks were considered the primary point of helmet contact in 37% of the bare helmet impacts and 67% of the padded helmet impacts.

The distribution of impact magnitudes for on-field impacts is displayed in Figure 8. Median (interquartile range) values for on-field bare helmet impacts were 23.61 (18.7 – 30.2) g PLA, 1349 (1047 – 1725) rad/$s^2$ PAA, 0.1256 (0.1028 – 0.1691) DAMAGE, and 2.37 (1.91 – 2.93) HARM. Median values for on-field padded helmet impacts were 24.21 (21.3 – 32.2) g PLA, 1099 (957 – 1849) rad/$s^2$ PAA, 0.1158 (0.0907 – 0.1502) DAMAGE, and 2.18 (1.76 – 2.79) HARM. Mann-Whitney test results revealed that there were no significant differences between the bare and padded helmet impacts in any metric (PLA: $p = 0.394$; PAA: $p = 0.245$; DAMAGE: $p = 0.093$; HARM: $p = 0.398$).

**Discussion:**

As concerns over sport-related brain injury risk continue to grow, innovations in head protective technologies are frequently being developed and brought to market. Padded helmet shell covers have gained in popularity across various levels of American football participation, with mandates on their use being implemented for specified training sessions at the elite level.[34] Though a number of previous studies have utilized laboratory experiments to evaluate the efficacy of various padded helmet shell covers,[2,7,49] to our knowledge the present study is the first published study to assess the efficacy of one of these products on the field, as well as investigate the padded helmet shell cover's response to changes in temperature and repeated impacts. Overall, we found that the padded helmet shell cover effectively attenuated angular head accelerations, DAMAGE, and HARM resulting from laboratory-controlled impacts, but reductions in linear head accelerations were inconsistent. Further, we found that the padded helmet shell cover exhibited a repeatable impact response when equipped to a helmet and only had small, but statistically significant, changes in impact attenuation across a wide range of temperatures. In a preliminary on-field investigation, we did not observe any significant reductions in any measure of impact severity after implementation of the padded helmet shell cover.

Padded protective headgears have been used for decades in a wide variety of contact sports. Laboratory tests of several padded headgears have supported their efficacy in reducing the magnitude of head kinematics resulting from impacts in sports such as rugby,[41] water polo,[10] and boxing.[32] Modern American football helmets are hard-shelled, and padded helmet shell covers represent the only currently available technology in the modern form of the sport that result in soft padding on the outside surface of an American football athlete's headgear. The padded helmet shell cover tested in this study loosely affixes to the exterior of a helmet's shell, claiming to attenuate impact magnitudes by adding an extra layer of foam padding that can compress axially, by sliding independently along the helmet shell at impact, and by providing an exterior surface friction that is lower than the helmet shell. In laboratory testing, only one of the helmets tested exhibited small, but statistically significant reductions in linear accelerations after implementation of the padded helmet shell covers. However, our study found that a significant reduction in angular acceleration was afforded by use of the padded helmet shell covers in all three helmets tested. Considering that angular accelerations are believed to contribute more to brain injury risk than linear accelerations,[20] these findings may be meaningful when assessing the potential clinical benefit of padded helmet shell covers. In line with these results, we found significant reductions in DAMAGE, a metric used to rapidly estimate maximum brain strain,[18] and HARM, a helmet performance metric found to be associated with on-field concussion incidence;[3,4] both metrics are derived, in part, from angular head kinematics. Our laboratory test results are consistent with previous studies that investigated other padded helmet shell cover devices, which found reductions in HARM[2] and angular head accelerations,[49] but limited efficacy in attenuating brain injury metrics derived from linear head accelerations.[7,49] Overall, our results in combination with the previous literature suggest that padded helmet shell covers are not particularly effective in attenuating linear acceleration from head impacts, but are effective in reducing angular accelerations in a laboratory setting.

Padded helmet shell covers are limited in their range of protection, in that they only cover the shell of the helmet, leaving the facemask unmodified. The facemask remains an important area for protection; facemask impacts can be particularly common on the field, as observed in our on-field dataset and previous literature.[2] To our knowledge, our study was the first to evaluate the effect of a padded helmet shell cover striking an opposing player's facemask. While we anticipate that a purely facemask-to-facemask impact would remain unchanged by addition of padded helmet shell covers, our data from laboratory facemask impacts revealed that head impact responses could be attenuated within a similar range as shell impact locations, despite only the padded helmet shell cover on the linear impactor's end cap being primarily engaged during the impact process.

Due to the sport's high rate of head impact exposure,[12,15] it is important that any helmet product used in American football maintains its ability to protect an athlete throughout its lifecycle of use. Testing of a cutout section of the padded helmet shell cover in drop testing suggested that after repeated impact exposure, the device's ability to attenuate force would deteriorate a small, but statistically significant amount. These findings are consistent with previous work that has found deterioration of impact attenuation provided by padded headgear and American football helmets after repeated, linear drop impacts.[14,24] Despite this, no association was found between HARM (a helmet performance metric found to be associated with on-field concussion incidence) and impact count when the padded helmet shell cover was affixed to a helmet and consecutively impacted 100 times. We believe that this, along with the lack of substantial reductions in peak linear accelerations in laboratory tests, suggests the primary mechanism of impact attenuation offered by the padded helmet shell cover is not linear force attenuation; rather, it is the low surface friction and independent movement from the helmet shell that consistently redirects impact energy throughout extended use. Similar technologies that decouple helmet layers exist in the interior of helmets used in other applications (e.g., cycling, snow sports, etc.) and have also been found to successfully attenuate angular head accelerations.[5,6,16,21]

An important, but relatively understudied aspect of helmet performance is the sensitivity of a given technology's impact response to changes in temperature. Existing helmet rating systems do not consider this in their rankings of helmet safety,[4,44] while standardized pass/fail testing overseen by NOCSAE only looks at helmets at ambient and hot conditions.[36] We found that significant variation in peak force attenuation could be observed between impacts to the padded helmet shell cover at various temperatures. Specifically, a 13% increase in peak force was found between the cold and hot temperatures conditions, which represented the extremes in which a helmet would likely be used. We considered this change in peak force to be relatively small compared to other, previously tested foams intended for use in American football helmets.[37] We did not perform full helmet testing across a range of temperatures, as the results would have likely been influenced by the sensitivity of the helmet's performance to changes in temperature; therefore, it remains unclear how these changes in peak force would translate to changes in brain injury risk. Nonetheless, further research is suggested to characterize the performance of helmets at varying temperatures to ensure optimal design of head protective technologies across different seasons and geographical locations.

Overall, results of our laboratory impact testing suggested significant reduction of angular head accelerations, DAMAGE, and HARM, along with consistent impact response after repeated impact exposure. However, no significant differences were observed in any measure of impact magnitude in our preliminary on-field investigation. One possible explanation for this is the lack of facemask coverage provided by the padded helmet shell covers. Over half of the impacts recorded in our on-field dataset of athletes wearing padded helmet shell covers involved both athletes' facemasks as a primary point of contact. As the padded helmet shell covers do not protect this area of a helmet, a purely facemask-to-facemask impact would remain essentially unchanged by implementation of the padded helmet shell cover. Therefore, it is possible that a large dataset consisting of only shell impacts would have yielded significant reductions in measures of angular acceleration, HARM, and brain strain similar to those observed in laboratory testing. However, our video resolution was not sufficient to determine if the facemask impacts in our study were purely facemask-to-facemask impacts without any contact with one or both padded helmet shell covers; therefore, we recommend that future studies carefully follow athletes with sufficient video recordings for detailed description of impact locations. Though not statistically significant, median PAA and median DAMAGE in the padded helmet shell cover dataset were lower than in the bare helmet dataset, so it remains possible that a larger sample size (i.e., recruitment of more participants and collection of more head impact events) would have yielded statistical significance. Part of the reason our sample size was smaller in this study was because we wanted to control for various sport-specific factors (i.e., player position, session type, impact mechanism) which have been found to influence the magnitude of head impact kinematics in American football.[9,30] For this study, we sought to eliminate the influence of these factors by analyzing kinematic data only from helmet-to-helmet impacts sustained by participants playing the linebacker position in practice sessions. However, our results could have been affected by differences in helmet models worn in the two years of study, which we did not track. Despite this possible explanation, previous research utilizing instrumented mouthguards has also found a lack of significant differences in head kinematics and brain strain between various American football helmet technologies,[9] so it cannot be discounted that the laboratory testing methodologies used to reconstruct helmet impacts may not sufficiently predict helmet performance in subconcussive American football impacts. Further, our lack of significant on-field findings may not be generalizable to all forms of padded helmet shell covers, all player positions, and skill levels in American football. For example, laboratory tests targeted specifically towards elite, professional lineman have suggested that padded helmet shell cover products other than the one tested in our study could be beneficial in mitigating head impact severity in that particular use case. However, that study (and no other previous study on padded helmet shell covers) has been validated on the field.

It is important to note that none of the on-field impacts recorded in our study resulted in a diagnosed concussion and we did not investigate the effect of the padded helmet shell covers on concussion incidence. Although previous studies have found that American football helmets with improved laboratory performance are associated with lower on-field concussion incidence,[3,13,39] several studies in other sports have found that the implementation of padded protective headgears did not yielded significant reductions in concussion incidence.[23,31,42] We speculate that while the padded helmet shell covers did not yield statistically significant reductions in measures of head accelerations, DAMAGE, or HARM for on-field subconcussive impacts, a reduction in concussion incidence could still be possible depending on level of competition, player position, and helmet model worn. Therefore, large-scale research studies that seek to determine the ability of padded helmet shell covers to reduce concussion incidence in various American football settings remain necessary.

The present study was not without limitations. While the laboratory equipment we utilized for full helmet impact tests is frequently used in injury biomechanics research, it may not perfectly reflect the true biomechanics of helmet-to-helmet impacts. The linear impactor, for example, translates only in a single axis. Our results may have differed if we used a second, helmeted headform for impact tests, which could also move in six degrees of freedom upon impact.[2] Padded helmet shell covers are often adopted by an entire team, such as the one we studied, and we therefore only studied the efficacy of the technology under the assumption that both athlete's equipped it to their helmets; it is unclear what effect we would have observed in our study if just one padded helmet shell cover was worn, such as has been investigated in previous studies.[2,49]

Further, the impact locations we tested were determined based on concussion data from the NFL.[4] The most frequent subconcussive impact locations have been found to differ from the impact locations most frequently associated with diagnosed concussions for at least one player position in football;[2] therefore, it may have been necessary to weight the effect of the padded helmet shell cover at each laboratory impact location based on its on-field frequency in order to better predict the on-field effect of equipping the technology. For repeated impact testing, we waited 80 seconds between impacts while the equipment and software readied itself for the next impact. However, repeated impacts in American football practice and competition could potentially occur at shorter time intervals, and it remains unclear how much our drop test or full helmet test results would have changed at shorter time intervals. For our preliminary on-field dataset, it is possible that several other factors, in addition to small sample size, could have had an influence on our results. For example, while we followed the same team in both years of this study, the team's approach or intensity in their practice sessions could have changed over time, therefore influencing the magnitude of the head impacts the athlete's sustained. Therefore, large-scale studies that acutely follow the same athletes with and without padded helmet shell covers equipped may uncover different findings from our study. Further, it should not be discounted that the addition of padded helmet shell covers could have a behavioral effect on athletes, which was not investigated in this study. Acute behavioral changes in athletes due to their change in head protection technology could not be monitored in this study due to the team adopting the headgear in 2020, when head impact monitoring was not possible due to the COVID-19 pandemic.

Overall, the laboratory results from this study suggest that padded helmet shell covers have the potential to reduce angular head accelerations and other brain injury risk metrics derived from angular head kinematics resulting from helmet impacts in American football. However, further research is needed to prove their efficacy in attenuating impact severity and injury incidence in a live, human setting. Future studies should investigate the effect of this relatively new technology in specific impact conditions (e.g. facemask vs. shell involvement) and at various levels of competition (e.g., high school and youth), as well as other sport and contact settings to determine its efficacy among different populations.

**FIGURES:**

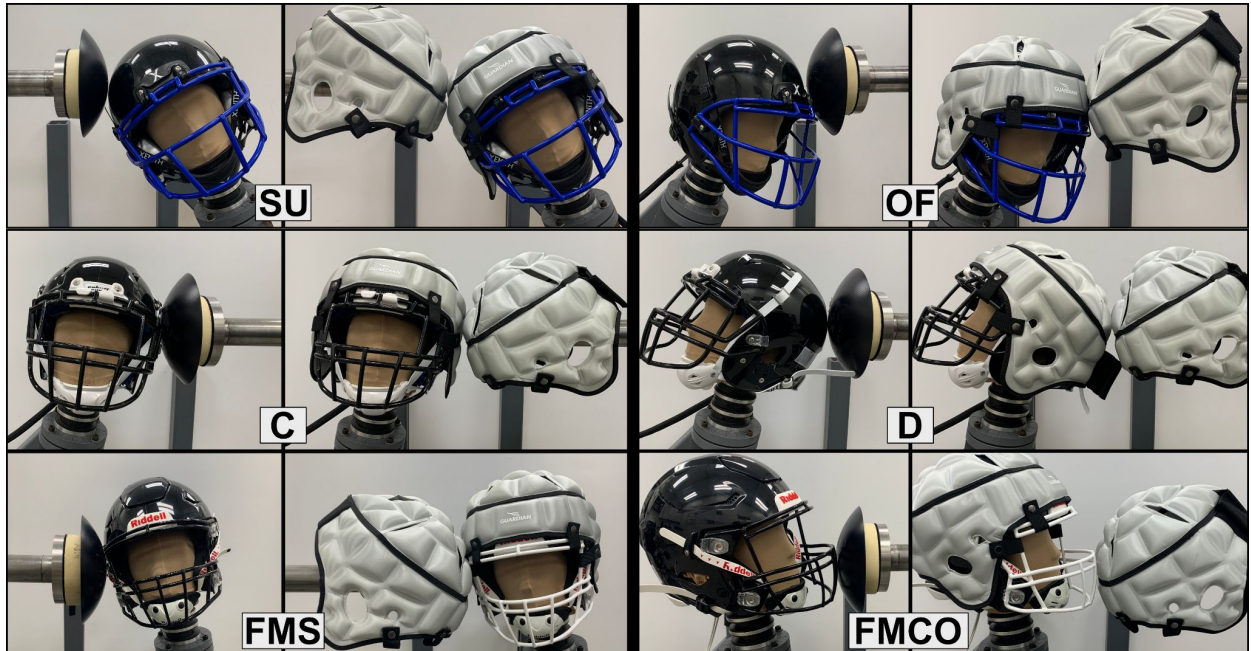

Figure 1: Helmet impact locations with and without the padded helmet shell cover affixed (SU = Side Upper; OF = Oblique Front; C = Side; D = Oblique Rear; FMS = Facemask Side; FMCO = Facemask Central Oblique). Examples of each helmet model tested are shown (top row = Xenith X2E+; middle row = Schutt Vengeance Pro; bottom row = Riddell Speedflex Diamond).

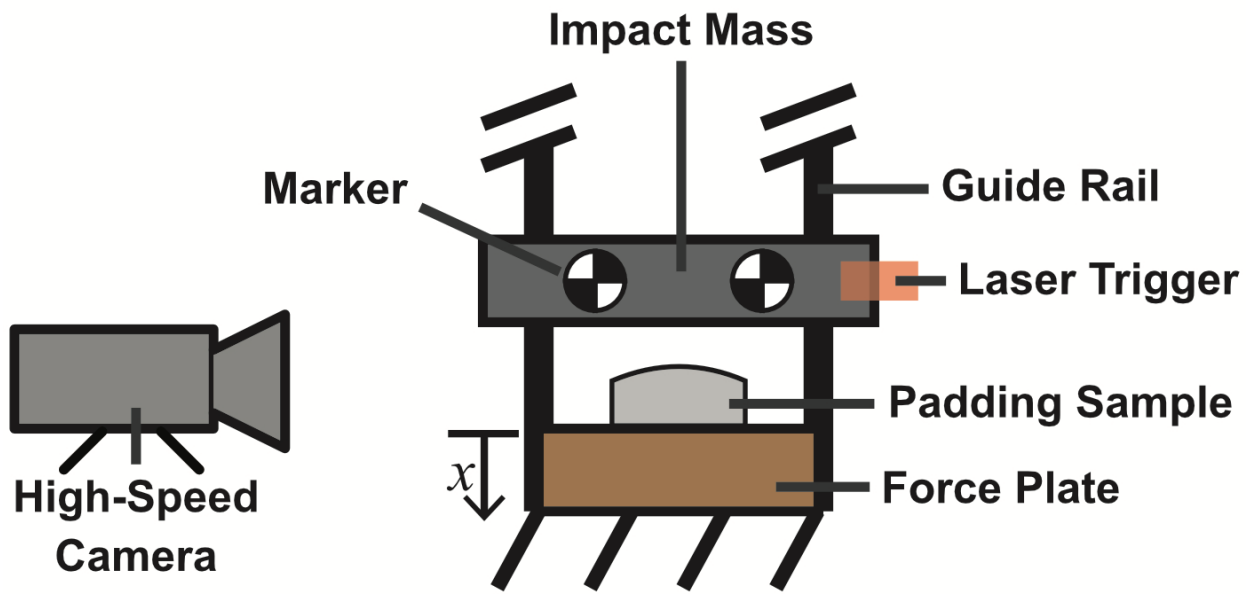

Figure 2: Laboratory setup for drop impacts performed on sample sections of the padded helmet shell cover.

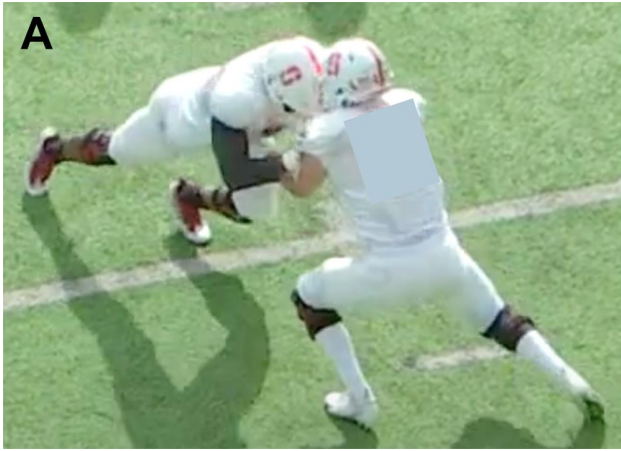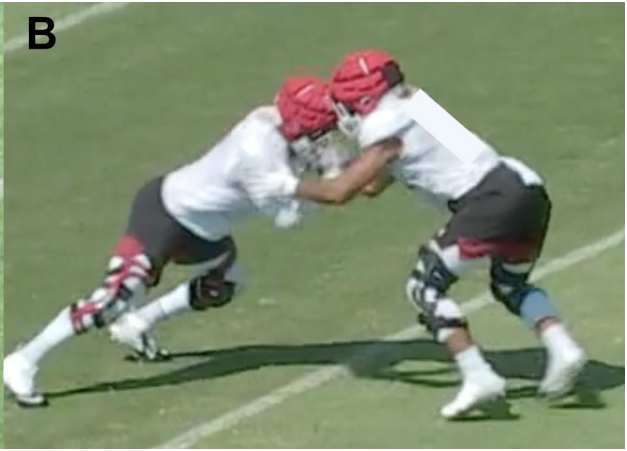

Figure 3: Still images of head impacts taken from videos of collegiate American football practices. Image (A) depicts an impact between bare helmets; image (B) depicts an impact between helmets equipped with padded helmet shell covers.

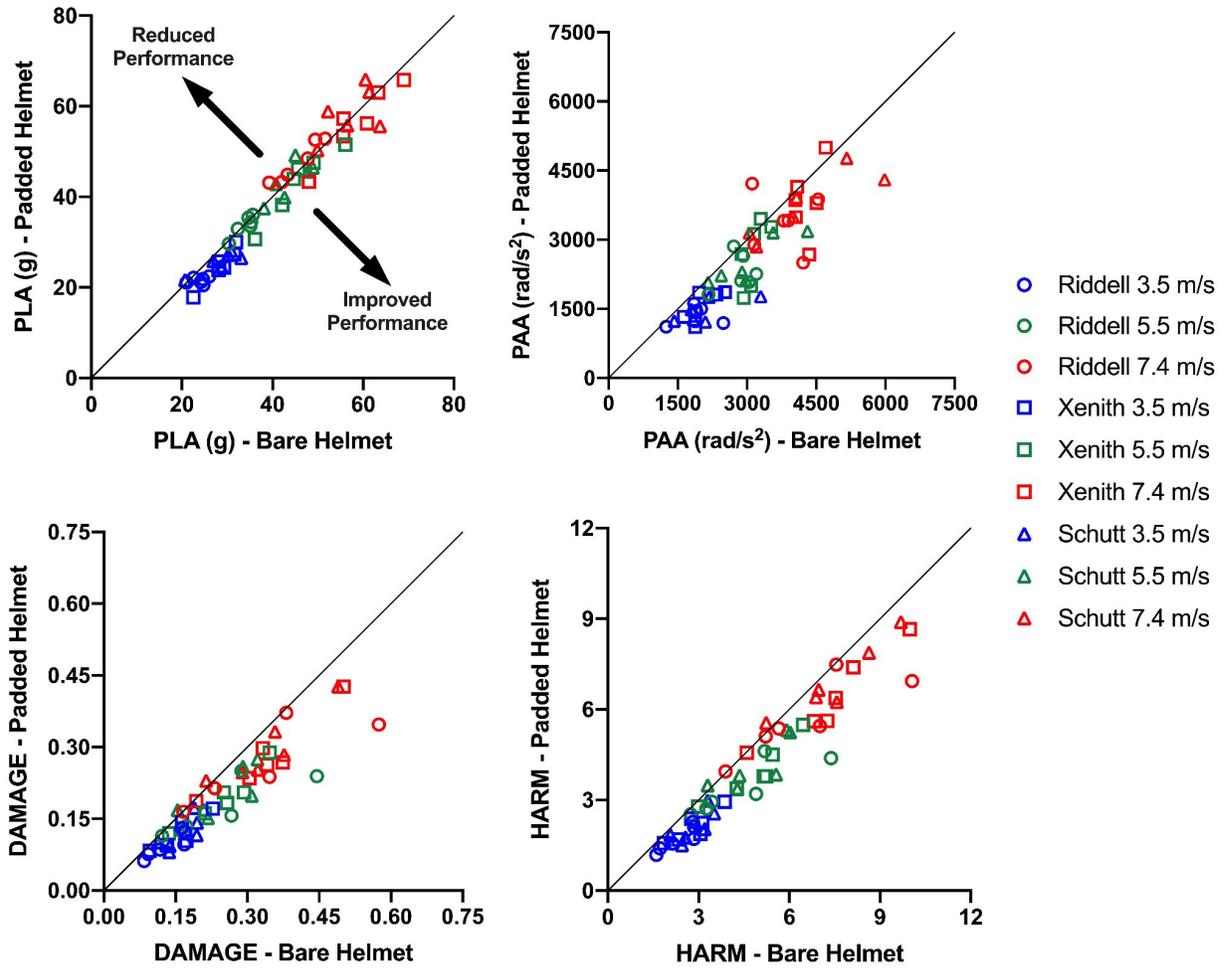

Figure 4. Comparison of helmet impact testing results between bare helmets and helmets equipped with the padded helmet shell cover. The black line represents the line of identity (slope = 1). Each point on the graph represents an average across three repeated impact trials. Area below and to the right of the line of identity indicates better performance with the padded helmet shell cover and area above and to the left of the line of identity indicates worse performance with the padded helmet shell cover.

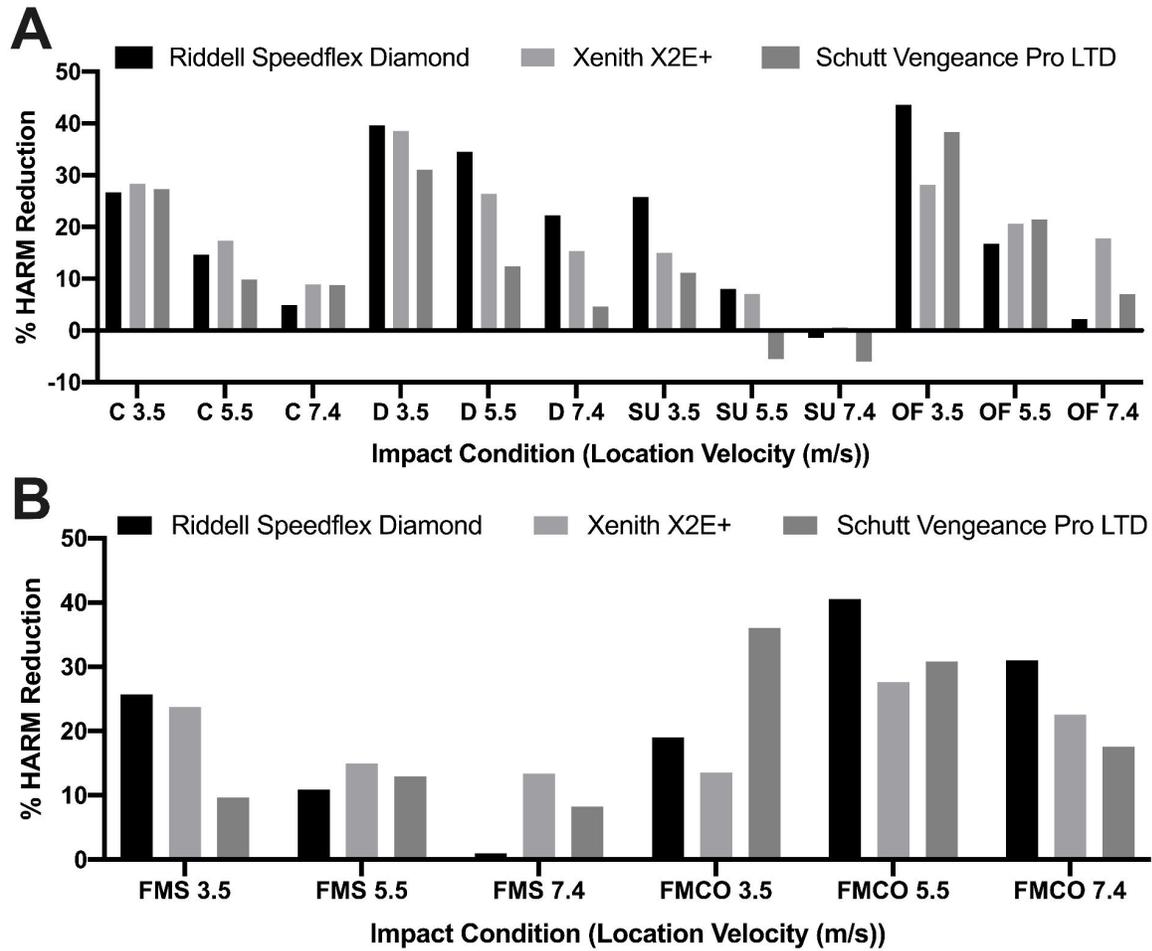

Figure 5: Percentage reduction in Head Acceleration Response Metric (HARM) provided by padded helmet shell covers for three helmets impacted at A) four helmet shell locations (C: Side; D: Oblique Rear; SU: Side Upper; OF: Oblique Front) and B) two facemask locations (FMS: Facemask Side; FMCO: Facemask Central Oblique).

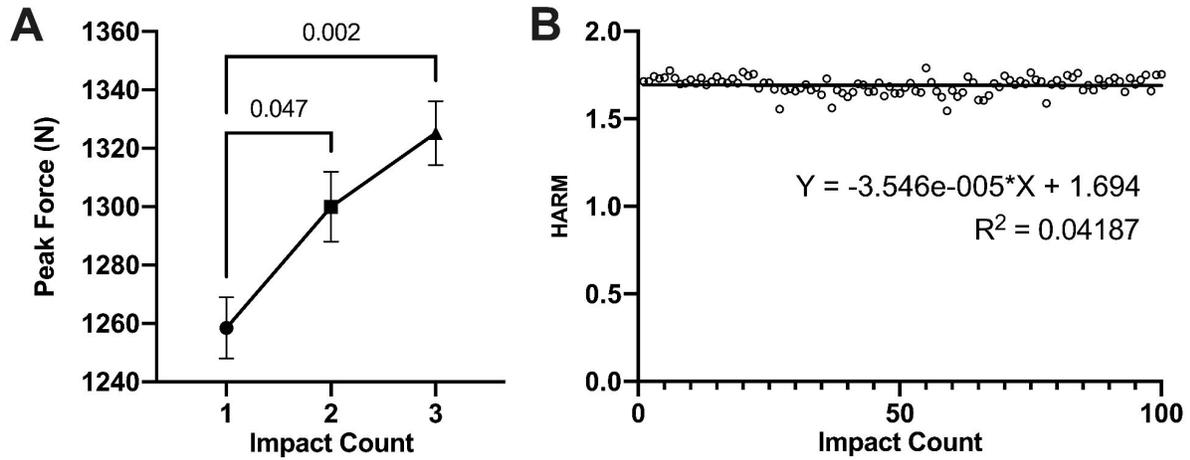

Figure 6: A) Peak force for repeated drop impact testing of a cutout section of the padded helmet shell cover. Mean values for each group of tests are shown, with error bars representing standard error of the mean. P-values below 0.1 from ANOVA are shown. B) HARM values for 100 consecutive impacts to the Oblique Front location of a Riddell Speedflex Diamond helmet equipped with a padded helmet shell cover. All impacts were delivered by a pneumatic linear impactor at 3.5 m/s, with its end cap covered with a padded helmet shell cover.

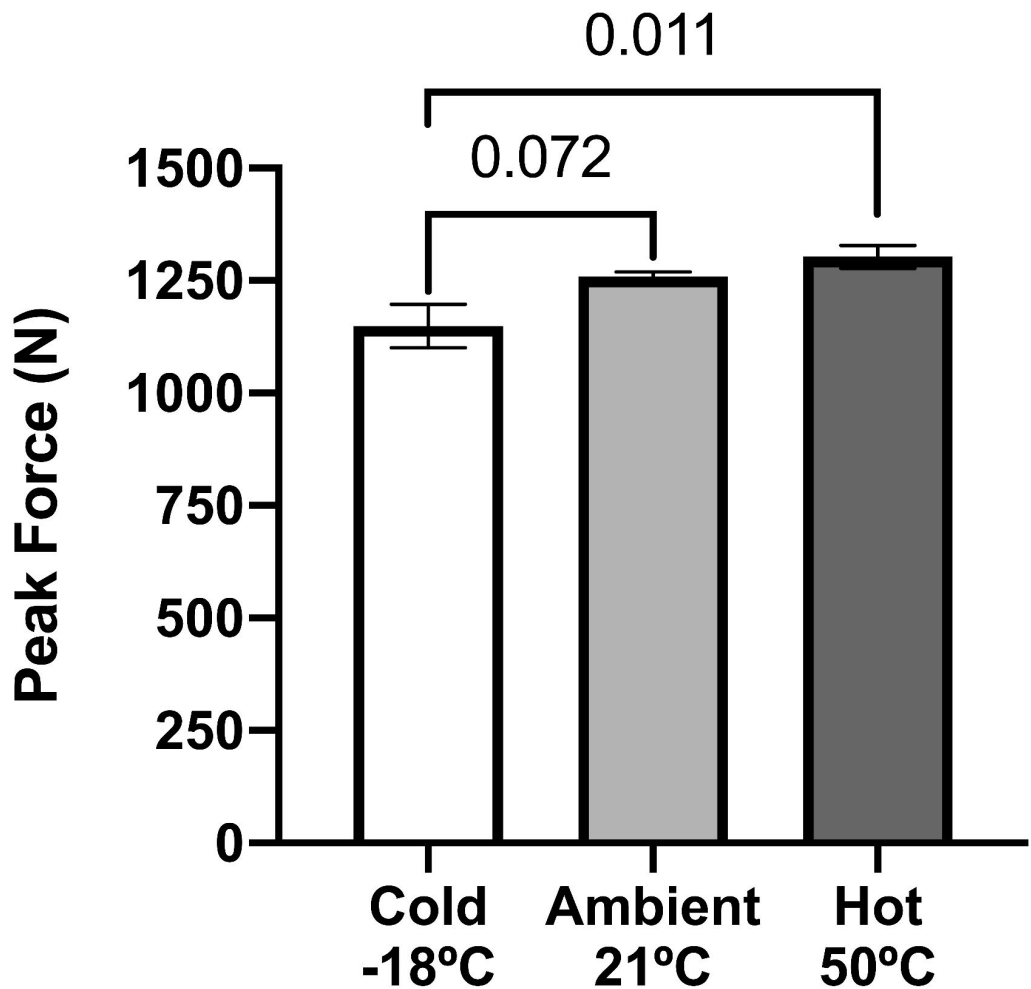

Figure 7: Peak force from impact testing on a cutout sample of the padded helmet shell cover. Each sample was exposed to the temperature condition for 18 hours prior to testing. Bar height represents the mean value of each group of tests with error bars representing standard error of the mean. P-values below 0.1 from ANOVA are shown.

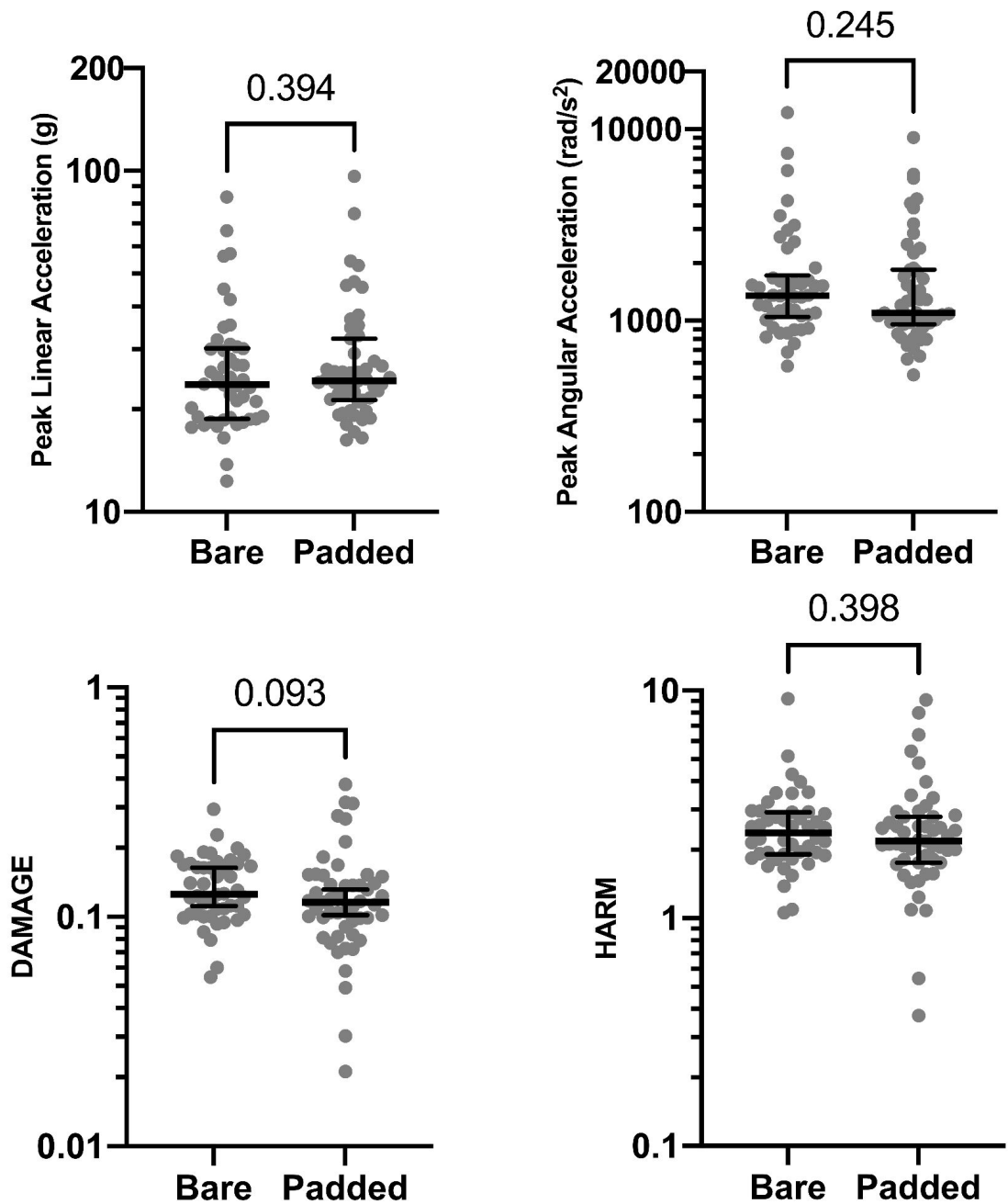

Figure 8: Distributions, displayed on logarithmic scale, of peak linear acceleration, peak angular acceleration, Diffuse Axonal Multi-Axis General Evaluation (DAMAGE), and Head Acceleration Response Metric (HARM), from helmet-to-helmet impacts between bare helmets (Bare) and helmets equipped with padded helmet shell covers (Padded). Middle lines represent the median values and error bars represent interquartile ranges.